\documentclass[useAMS,usenatbib]{mn2e}

\usepackage{amssymb,amsmath,graphicx,epsfig}

\title[The H{\small I} environment of counter-rotators]{
       The H{\sc i} Environment of Counter-rotating Gas Hosts:\\
       Gas Accretion from Cold Gas Blobs}
\author[Chung et al.]{Aeree Chung$^{1}$\thanks{E-mail:achung@yonsei.ac.kr},
        Martin Bureau$^{2}$\thanks{E-mail:bureau@astro.ox.ac.uk},
        J. H. van Gorkom$^{3}$\thanks{E-mail:jvangork@astro.columbia.edu}, and
        B{\"{a}}rbel Koribalski$^{4}$\thanks{E-mail:Baerbel.Koribalski@csiro.au}\\
$^{1}$Department of Astronomy and Yonsei University Observatory, Yonsei University, Seoul 120-749, Korea\\
$^{2}$Sub-Department of Astrophysics, University of Oxford, 
      Denys Wilkinson Building, Keble Road, Oxford OX1 3RH, United Kingdom\\
$^{3}$Department of Astronomy, Columbia University, 550 West 120th Street, New York, NY 10027, U.S.A.\\
$^{4}$Australia Telescope National Facility, CSIRO, PO Box 76, Epping, NSW 1710, Australia}

\begin{document}

\date{Draft Version}

\pagerange{\pageref{firstpage}--\pageref{lastpage}} \pubyear{2002}

\maketitle

\label{firstpage}

\begin{abstract}
We probe the H{\sc i} properties and the gas environments of three early-type barred
galaxies harbouring counter-rotating ionized gas, NGC~128, NGC~3203 and NGC~7332. 
Each system has
one or more optically-identified galaxy, at a similar or as yet unknown redshift
within a 50 kpc projected radius. Using H{\sc i} synthesis imaging data, we investigate
the hypothesis that the counter-rotating gas in these galaxies has been accreted from their
neighbours. In NGC~128 and NGC~3203, we find $9.6\times10^7$ and 
$2.3\times10^8$~$M_\odot$ of H{\sc i}, respectively, covering almost the entire stellar
bodies of dwarf companions that appear physically connected. Both the H{\sc i} morphology 
and kinematics are suggestive of tidal interactions.
In NGC~7332, we do not find any directly-associated H{\sc i}. Instead, NGC~7339, 
a neighbour of a comparable size at about 10~kpc, is found with 8.9$\times10^8~M_\odot$ of
H{\sc i} gas. 
More recently in a single dish observation, however, another group discovered a 
large H{\sc i} structure which seems to be an extension of NGC~7339's H{\sc i} disc and also
covers NGC~7332. 
All these observations thus suggest that H{\sc i} gas is being accreted in these three
galaxies from their companions, which is likely responsible for the kinematically-decoupled
gas component present in their central region.
In particular, the dynamical friction timescales of the nearest neighbours with H{\sc i} gas
of NGC~128 and NGC~3203
are comparable to their orbital timescales around the counter-rotators, several $\sim10^8$ yr,
implying that those neighbours will likely soon merge with the primary galaxies,
fueling them with gas. NGC~7332 also appears to be in the merging process with its
neighbour through the common H{\sc i} envelope.
Besides, we find some other potential gas donors around NGC~128 and NGC~7332: 
two H{\sc i}-rich galaxies with $M_{\rm HI}=3.8\times10^8$ and $2.5\times10^9~M_\odot$ 
at a distance of $\approx$67 kpc from NGC~128, and two dwarf systems
with $M_{\rm H{\small I}}=3.9\times10^7$ and $7.4\times10^7~M_\odot$ 
at $\lesssim$100~kpc from NGC~7332.
Among the seven H{\sc i} features identified in this study, three of them are 
associated with dwarf galaxies,
two of which have only been recently identified in a blind survey
while the third one is still not catalogued at optical wavelengths.
Considering the incompleteness of existing studies of the faint dwarf galaxy population 
both in the optical and in H{\sc i}, accretion from cold gas blobs,
presumably gas-rich dwarfs, is expected to occur even more frequently than what is 
inferred from such cases that have been observed to date.
\end{abstract}

\begin{keywords}
ISM: kinematics and dynamics -- galaxies: bulges -- galaxies: evolution
-- galaxies: interactions -- galaxies: structure

\end{keywords}

\section{Introduction}
\label{intro}
Morphological and dynamical peculiarities of galaxies can be used as diagnostics of galaxy 
formation. Features such as tails, bridges, rings and distinct kinematic or
stellar population components help us to better understand the structural
parameters of galaxies and their formation mechanisms
\citep[e.g.][]{hb91,bekki98,bertola98}.

A counter-rotating gas disc during galaxy formation is dynamically 
unstable. Kinematically-distinct gas components are therefore more likely to
be the result of subsequent accretion than of being primordial \citep{pp01}.
Counter-rotating gas is more common in elliptical and lenticular (S0) galaxies
than gas-rich spirals (see Bureau \& Chung 2006 for an extensive discussion
of the frequency in S0s; Sarzi et al. 2006 and Davis et al. 2011, for ellipticals).
As \citet{kf01} argue, this probably reflects both that 1) it
is harder for counter-rotating gas to survive in late-type galaxies due to friction
within co-rotating material, and 2) the accretion event itself can
help consume most of the gas in a triggered starburst, which may
yield a gas-poor early-type system with some kinematically-distinct gas leftover. Hence studying the origin(s) of counter-rotating gas
is important to understand the formation and evolution of galaxies,
in particular that of the early-type population, and growth through
gas accretion and minor mergers \citep[][]{bjc07,gb10}.

For the origin of counter-rotating gas, sporadic or continuous gas
infall from the galactic halo and merging with gas-rich systems have been 
suggested \citep[see][and references therein]{tr98}. Alternatively, gas accretions
from filaments, predicted to occur frequently in low-density regions today
\citep[e.g.][]{keres05}, may well form kinematically-distinct components. 
As some simulations have shown
\citep[e.g.][]{tr96}, continuous gas infall might be best suited for the
creation of extensive counter-rotating gas discs. Observationally, on the other hand, 
a number of counter-rotators with hints of minor merging events have
been found \citep[e.g.][]{haynes00}, although we still lack bullet-proof
evidence of gas accretion directly associated with these merging events.
One such $smoking~gun$ example is the galaxy pair NGC~1596 and NGC~1602
\citep{chung06}.

NGC~1596 is one of the counter-rotators serendipitously found by \citet{cb04}
in the sample of \citet{bf99}. As shown in Figure 1 of \citet{bc06}, the galaxy 
hosts ionized gas that in the central few kpc is rotating in the opposite direction to the bulk of 
the stars. The counter-rotating gas is kinematically
asymmetric, extending up to 150~km~s$^{-1}$ on one side of the disc while
it extends less than 50~km~s$^{-1}$ on the other side.
There is a companion with a similar systemic 
velocity, NGC~1602, at distance of 20~kpc (from centre to centre).
A deep optical image by \citet{pohlen04} revealed extended stellar envelopes
around both galaxies, with no clear border in-between
\citep[see Figure 1 in][]{chung06}. In an H{\sc i} follow-up study using the 
Australia Telescope Compact Array (ATCA), 
\citet{chung06} found a large H{\sc i} structure covering both galaxies. The H{\sc i} 
morphology and kinematics are quite regular with no sign of disturbance
within the stellar disc of NGC~1602, while they become highly asymmetric in the
outer disc. An H{\sc i} tail can clearly be seen on one side, leading from 
NGC~1602 to the centre of NGC~1596. This strongly supports
the hypothesis that much of the gas in NGC~1596 originates from its gas-rich companion 
and must have been heated while entering NGC~1596, as it is observed
in the form of ionized gas in NGC~1596.

Besides NGC~1596, \citet{cb04} identified two more counter-rotators, 
NGC~128 and NGC~3203 \citep[see also][]{ea97,bc06}. As shown in Fig. 1
of \citet{bc06}, the counter-rotating ionized gas in these galaxies is found to be
a few kpc in size with a maximum rotation velocity comparable to the stars
($\approx150~$km~s$^{-1}$). The counter-rotating component is quite
symmetric in the case of NGC~128, being widely spread to both sides of 
the disc, while it is mostly found on one side of the disc in NGC~3203.
Both systems have closeby 
galaxies at similar velocities or candidate companions that are potential
donors of counter-rotating gas, like NGC~1602. In this work we
present results from our H{\sc i} follow-up study of these galaxies, 
using the Very Large Array (VLA)\footnote[1]{The National Radio 
Astronomy Observatory is a facility of the National Science Foundation 
operated under cooperative agreement by Associated Universities, Inc.}
to study their H{\sc i} properties and environment.
In addition, we have included another well-known counter-rotator 
of similar morphology, the barred S0 galaxy NGC~7332 from the study of 
\citet[][]{fb04}. 

In Section 2, we describe the general properties and the environment of our
sample galaxies. In Section 3, we summarize the H{\sc i} observations and data reduction.
In Section 4, we describe the H{\sc i} properties of the sample galaxies and 
their surroundings.
In Section 5, we discuss the origin of the counter-rotating ionized gas and
its role in galaxy evolution. We conclude in Section 6.
We adopt a distance of 53.9~Mpc and 38.2~Mpc to NGC~128 and NGC~3203, respectively,
based on the standard cosmology with $H_0=$73~km~s$^{-1}$~Mpc$^{-1}$, 
$\Omega_M=$0.27 and $\Omega_\Lambda=$0.73 \citep{cosmos03}.
For NGC~7332, we adopt 23~Mpc, a distance determined using the surface brightness
fluctuation technique \citep{tonry01}, which is probably the most reliable distance for this system.

\begin{figure*}
\includegraphics[width=18cm]{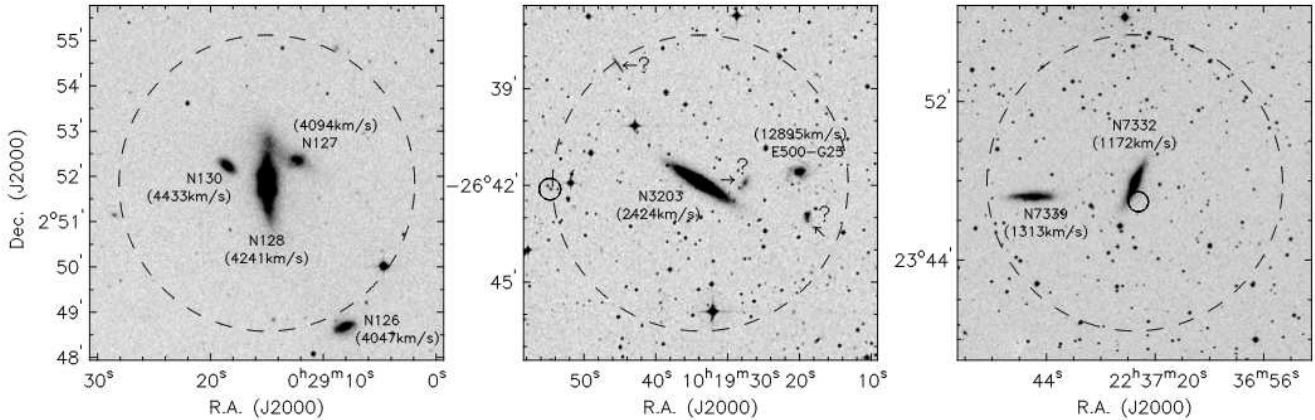}
\caption[]{$B$-band optical images of the three counter-rotating gas hosts from 
the second Digitized Sky Survey (DSS2). From left to right: NGC~128, NGC~3203 and
NGC~7332. The large dashed circles are drawn at 
a 50 kpc projected radius from the counter-rotators. The sample galaxies and their neighbours 
at $<$50 kpc radii with a known redshift are labeled, with the optical velocity. Small 
solid circles represent catalogued galaxies with no redshift within a $<50$~kpc radius.
All the optically-catalogued extragalactic sources (with or without 
a redshift) are listed in Table~\ref{tbl-sample}. Extended sources within this 
radius that are not cataloged are indicated with an arrow. \label{fig-sample}}
\end{figure*}

\section{Sample}
\label{sample}
Our sample consists of three edge-on disc galaxies, NGC~128, NGC~3203
and NGC~7332, where ionized gas ([O{\sc iii}], but also H$\beta$ for NGC~7332) has
been found to rotate in the opposite direction to the stars 
\citep{ea97,cb04,fb04,bc06}. Besides hosting kinematically-decoupled ionized gas 
in their centre, the galaxies in the sample have other
properties in common. 

Firstly, each galaxy is surrounded by several definite or potential 
companions. Figure~\ref{fig-sample} shows optical images of the galaxy  
environments. Dashed circles are drawn at 50~kpc projected radii around 
the counter-rotators at the adopted distance for each galaxy
(angular scales of 15.3, 10.9 and 6.69~kpc per arcminute, respectively). Within 50~kpc radii, 
the redshifts of relatively bright neighbours (at most 15 times fainter than
the counter-rotator, i.e. $\Delta m_B\lesssim3$) are
always available, and their group membership is well defined. That is, 
NGC~127 and NGC~130 near NGC~128, and NGC~7339 near NGC~7332, are
likely true neighbours, while ESO~500-G023 in NGC~3203's field must
be a background object. The redshifts of dwarf galaxies fainter than that 
are not available and their association with the counter-rotators remains 
an open question. These optically-visible but
small systems within the dashed circles (whether catalogued or not)
are indicated with question marks in Figure~\ref{fig-sample}. 
At optical wavelengths, only one galaxy among the sample, NGC~128, shows 
direct evidence for tidal interaction with its neighbour NGC~127, as
clearly indicated by the stellar tail between them.

Secondly, all galaxies in the sample have a boxy or peanut-shaped (B/PS)
bulge, which is likely the edge-on view of a 
thickened bar (see Bureau \& Freeman 1999 and
Bureau \& Athanassoula 2005, and references therein).
In fact, the presence of a bar has been kinematically confirmed in 
NGC~128 and NGC~3203 in the stellar kinematic study of \citet{cb04},
and in NGC~7332 by \citet{fb04}. The relevance of the counter-rotating
gas and the presence of the bars will be discussed in more detail
in Section~\ref{discuss}.

The general properties of the sample galaxies and their neighbours are
summarized in Table~\ref{tbl-sample}. 


\begin{table*}
\centering
\begin{minipage}{140mm}
\caption{General properties and environment of the sample galaxies$^{\rm a}$.\label{tbl-sample}}
\begin{tabular}{lllrllrllr}
\hline \hline & 
\multicolumn{3}{c}{NGC~128}  & 
\multicolumn{3}{c}{NGC~3203} & 
\multicolumn{3}{c}{NGC~7332}\\ \hline
Coordinate: &&&&&&&&&\\
~~~$\alpha_{2000}$ ($~~^h~~^m~~^s$) & 
\multicolumn{3}{c}{00 29 15.0} & 
\multicolumn{3}{c}{10 19 34.5} & 
\multicolumn{3}{c}{22 37 24.5}\\
~~~$\delta_{2000}$ ($~~^\circ~~~'~~~''$) & 
\multicolumn{3}{c}{ 02 51 55.0}& 
\multicolumn{3}{c}{-26 41 53.0}&
\multicolumn{3}{c}{ 23 47 52.0}\\
Morphology & 
\multicolumn{3}{c}{S0 pec sp} &
\multicolumn{3}{c}{ SA(r)0$+?$ sp} &
\multicolumn{3}{c}{ S0 pec sp}\\
$D_{25}$ (arcmin)$^{\rm b}$&
\multicolumn{3}{c}{ 3.0}&
\multicolumn{3}{c}{ 2.9}&
\multicolumn{3}{c}{ 4.1}\\
$B_{\rm T}$ (mag)$^{\rm c}$&
\multicolumn{3}{c}{12.77}&
\multicolumn{3}{c}{13.10}&
\multicolumn{3}{c}{12.02}\\
$V_{\rm opt}$ (km~s$^{-1}$)$^{\rm d}$& 
\multicolumn{3}{c}{4241}& 
\multicolumn{3}{c}{2394}& 
\multicolumn{3}{c}{1172}\\
Neighbours$^{\rm e}$ &\multicolumn{9}{c}{-------- Object name ($\Delta d$ in arcmin,
  $\Delta V_{\rm opt}$ in km~s$^{-1}$) --------}\\
&N127 &(0.8, &-147)& J10195473$^{f}$ &(4.7, &N/A)& N7339 &(5.2, &141)\\
&N130 &(1.0, & 192)& & & & J22372366$^{f}$ &(1.0, &N/A)\\
\hline
\end{tabular}
\tiny
$^{a}${The information about the three sample galaxies has been obtained from the Third
Reference Catalog of Bright Galaxies \citep[RC3;][]{rc3}, while the data on their neighbours
have been collected from the NASA/IPAC Extragalactic Database (NED; http://nedwww.ipac.caltech.edu/).}\\
$^{b}${Mean apparent major isophotal diameter, measured at or reduced to the surface brightness 
level $\mu_B$ = 25.0 $B$ mag~arcsec$^{-2}$.}\\
$^{c}${Total (asymptotic) magnitude in $B$, derived by extrapolation from 
photoelectric aperture and surface photometry with photoelectric 
zero point data.}\\
$^{d}${Mean heliocentric radial velocity, derived from optical observations.}\\
$^{e}${Optically identified neighbours within a 50~kpc projected radius and 
  $\Delta V_{\rm opt}\lesssim$1000~km~s$^{-1}$ or no redshift. $\Delta d$ is
the projected distance.}\\
$^{f}${2 Micron All Sky Survey \citep[2MASS;][]{2mass}.}\\
\end{minipage}
\end{table*}

\begin{table*}
\centering
\begin{minipage}{100mm}
\caption{VLA observing parameters.\label{tbl-vlaobs}}
\begin{tabular}{lccc}
\hline \hline& NGC~128 & NGC~3203 & NGC~7332\\ \hline
Phase centre: & & & \\
~~~~~~~~$\alpha_{2000}$ ($~~^h~~^m~~^s$) & 00 29 15.1 & ~10 19 34.4 & 22 37 35.9\\
~~~~~~~~$\delta_{2000}$ ($~~^\circ~~~'~~~''$)& 02 51 50.0& -26 41 54.0& 23 47 32.0\\
Configurations & D & DnC, D & C, DnC\\
Velocity centre (km~s$^{-1}$)$^{\rm a}$ & 4241 & 2409 & 1375\\
Bandwidth (MHz) & 6.25 & 3.125 & 3.125\\
Channel width (km~s$^{-1}$) & 41.6 & 10.4 & 10.4 \\
Synthesized beam FWHM ($''$) & 56.9$\times$50.3 & 48.7$\times$35.6$^{\rm b}$ & 18.8$\times$16.5$^{\rm b}$\\
Noise (1$\sigma$)$^{\rm c}$: & & & \\
~~~~~~~~in mJy beam$^{-1}$ & 0.28 & 0.60 & 0.33\\
~~~~~~~~in 10$^{19}$ cm$^{-2}$ & 0.45 & 0.40 & 1.23\\
\hline
\end{tabular}
\tiny
$^{a}${Heliocentric velocity using optical definition.}\\
$^{b}${Combined data from all configurations.}\\
$^{c}${Rms of the combined cube per channel. See text for further details.}
\end{minipage}
\end{table*}

\begin{figure*}
\includegraphics[height=6cm]{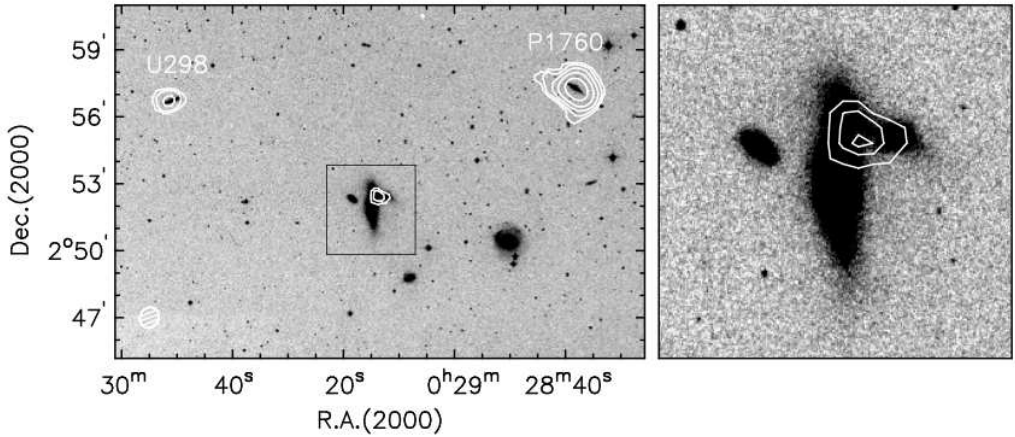}
\caption[]{On the left, the H{\sc i} distributions of NGC~128 and its surrounding
are shown in white contours, overlaid on a grayscale optical blue image from
second Digital Sky Survey DSS2. The H{\sc i} contours are 2, 4, 8,
..., $\times10^{19}$~cm$^{-2}$. The synthesized beam is shown at the bottom
left. The H{\sc i} gas to the northeast of NGC~128, that is likely bound to
UGC~298, appears to be disturbed, although no direct evidence of interaction with NGC~128 is found.
The H{\sc i} to the northwest of NGC~128, that nicely coincides with PGC~001760,
also does not show any sign of interaction.
The outlined region of $4'\times4'$ around NGC~128 is magnified on
the right. The contours are drawn at H{\sc i} column densities of 
2, 4 and 6$\times10^{19}$~cm$^{-2}$. Some H{\sc i} gas is found to extend over the two
galaxies NGC~127 and NGC~128 (a centre to centre projected distance of $\approx15~$kpc). 
The H{\sc i} morphology is irregular, with the peak flux between the two galaxies. The H{\sc i} 
line profiles are presented in Figure~\ref{fig-n0128gp}.\label{fig-n0128hi}}
\end{figure*}

\begin{figure*}
\includegraphics[bb=40 195 560 380,height=5cm]{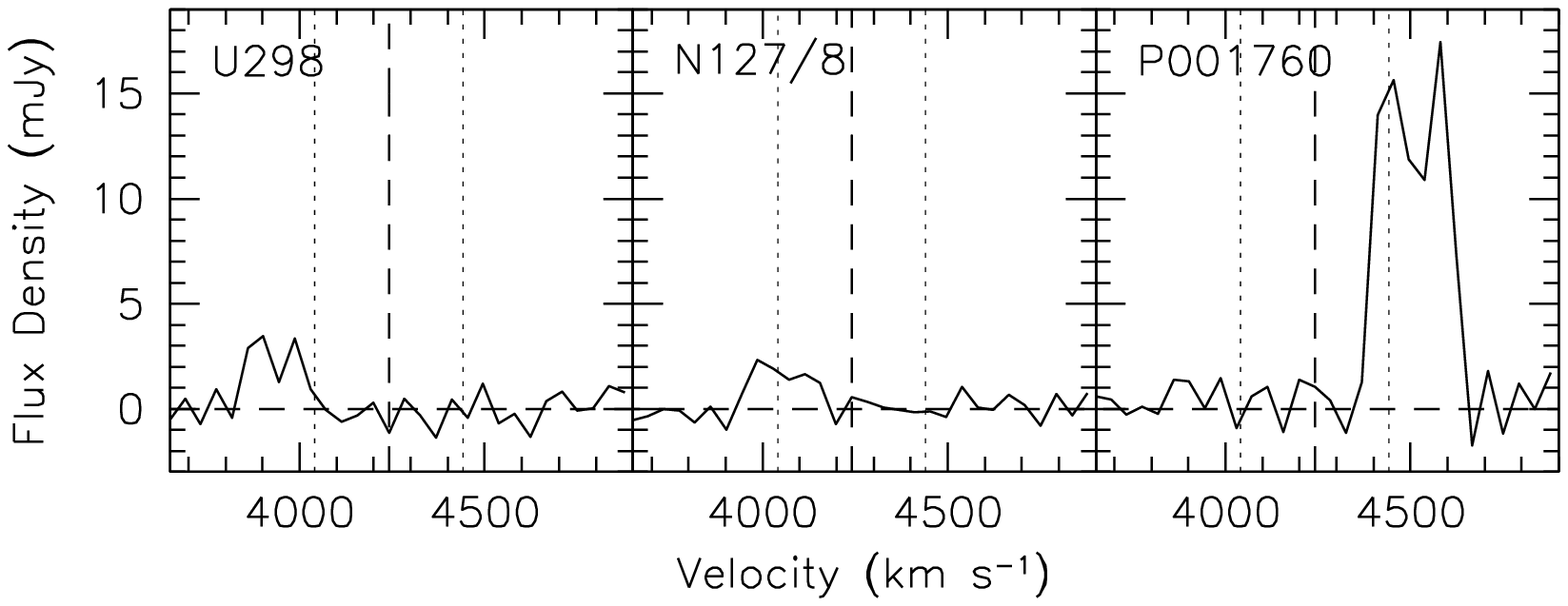}
\caption[]{Total line profiles of the H{\sc i} structures found in NGC~128's field.
The dashed line in each panel represents the optical velocity of NGC~128;
the velocity range covered by NGC~128's rotation ($V_{\rm rot}\approx200$ km~s$^{-1}$) 
is indicated by the dotted lines \citep{cb04}.
The H{\sc i} gas found to the northeast coincides well with UGC~298 and is likely
to be associated with it, although no prior redshift measurement is
available. The velocity of the H{\sc i} gas to the northwest of the
field coincides well with the optical redshift of PGC~001760. The H{\sc i} gas found between
NGC~127 and NGC~128 is spatially offset from both galaxies, although its 
velocity range is close to the optical velocity of NGC~127
(4094~km~s$^{-1}$). 
This H{\sc i} gas is likely to have been originally bound to NGC~127 and is
now being transferred to NGC~128. 
\label{fig-n0128gp}}
\end{figure*}

\section{Observations \& Data Reduction}
\label{obs}
The VLA observations were done in 2004-5 in several configurations. NGC~128
and NGC~3203 had only H{\sc i} flux upper limits at the time of
observing \citep[1.36 and 6.50~Jy~km~s$^{-1}$ for NGC~128 and NGC~3203, respectively;][]{hr89}
and were thus observed in D array - the most sensitive array of the VLA. NGC~3203,
located in the southern sky ($\delta<-26~$deg), was also observed with the DnC 
hybrid configuration to obtain a better coverage of the $u-v$ plane. NGC~7332, with a 
prior H{\sc i} flux measurement \citep[0.84~Jy~km~s$^{-1}$;][hereafter RC3]{rc3}, 
was observed in the C and DnC configurations. 

Each galaxy was observed with 3.125 and 6.25~MHz bandwidths at the same time. The 
narrow band covers roughly $660$~km~s$^{-1}$, providing good coverage of the host galaxy,
while the wide band covers $>1300$~km~s$^{-1}$, probing a broader velocity 
range across the sample galaxy fields. The correlators were configured to 
produce 127 and 63 channels with a single polarization, respectively. 
Online Hanning smoothing was applied and every other channel was discarded. 
This resulted in 63 and 31 independent channels with a velocity resolution of 10.4 
and 41.6~km~s$^{-1}$, respectively.

The data reduction was done using the Astronomical Imaging Processing System ({\tt\small AIPS}).
Data from different configurations and bandwidths were calibrated separately. 
For NGC~3203 and NGC~7332, the visibility data of different baselines were merged using the
AIPS command {\tt DBCON} after flux, phase and bandpass calibration, while 
the data with different velocity resolutions were treated separately. 
The continuum was subtracted using a linear
fit to the $u-v$ data of line-free channels at both sides of the band. 
Larger ranges were used for the 6.25~MHz bandwidth data. High $u-v$ points caused by 
interference were flagged after continuum subtraction. We first made cubes of a 
large field of view with a range of weighting, to search for H{\sc i} emission from sources 
far away from the field centre. The final cubes were then made using ROBUST=1 
\citep{briggs95} to maximize sensitivity while keeping a good spatial resolution. 

\begin{figure}
\includegraphics[height=12cm]{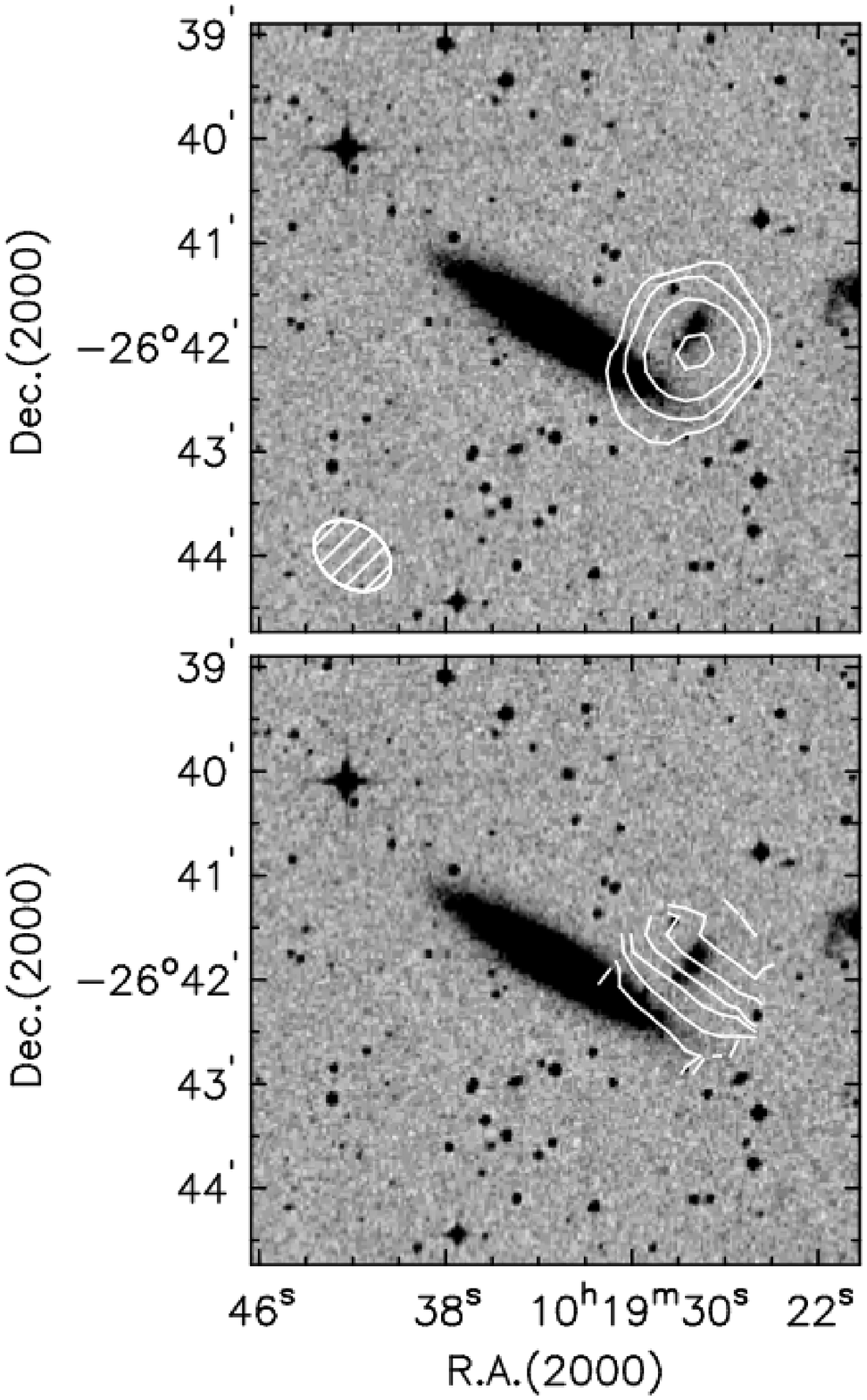}
\caption[]{Top: H{\sc i} distribution of NGC~3203 and its surroundings, shown in 
white contours on top of the DSS2 blue image. The synthesized beam is
shown at the bottom left. The H{\sc i} contours are 3, 6, 12 and
24$\times10^{19}~$cm$^{-2}$. An H{\sc i}-rich dwarf galaxy is found near NGC~3203,
not catalogued at any wavelength yet. Although the peak
of the H{\sc i} gas is slightly offset from the optical centre of the dwarf
($\approx 6''$ or 1~kpc), the optical and H{\sc i} emission more or less coincide on the sky,
with almost the same major axis (perpendicular to NGC~3203's disc).
Bottom: H{\sc i} velocity field. Starting from the contour crossing NGC~3203's disc,
the velocity contours are 2520, 2530, 2540, ..., and 2580 km~s$^{-1}$.
It is clearly seen that the iso-velocity contours are almost prependicular to 
the major axis of the dwarf companion.
This strongly suggests that the H{\sc i} cloud is rotationally supported and bound to the dwarf.
\label{fig-n3203hi}}
\end{figure}

\begin{figure}
\includegraphics[bb=40 195 280 380,height=6cm]{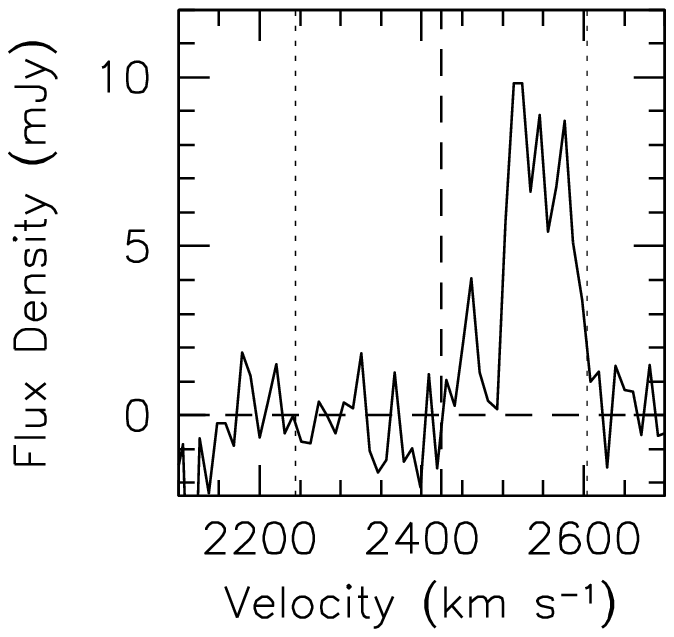}
\caption[]{Total H{\sc i} line profile of the gas structure shown in 
Figure~\ref{fig-n3203hi} (AC dwarf I). The dashed line represents the optical
velocity of NGC~3203; the dotted lines indicate the rotational velocity
range ($\approx180$ km~s$^{-1}$), taken from \citet{cb04}. The $>3\sigma$ feature close to NGC~3203's
systemic velocity is suggestive of gas being transferred from the 
dwarf to NGC~3203. The H{\sc i} cloud also morphologically appears to be 
more extended toward NGC~3203's disc, as seen in Figure ~\ref{fig-n3203hi}.
\label{fig-n3203gp}}
\end{figure}

The total H{\sc i} images were made by taking moments along the frequency
axis (0th moments). The task {\tt MOMNT} creates a mask to blank the 
images at a given cutoff level. In creating the masks, we applied Gaussian 
and Hanning smoothing spatially and spectrally, respectively, to maximize the 
signal-to-noise. 

Two sets of cleaned cubes per object were produced,
one set for the narrow band data and one for the wide band. For each galaxy, 
the two sets were inspected separately and only the set showing the H{\sc i} 
features better is presented in this paper: 6.25~MHz data for 
NGC~128 and 3.125~MHz data for NGC~3203 and NGC~7332.
The VLA observing parameters, the spectral and spatial resolutions, and the
rms of the final cubes used to produce the H{\sc i} maps presented in this work
are summarized in Table~\ref{tbl-vlaobs}.
The total H{\sc i} mass in solar unit has been calculated using the formula
$M_{\rm H{\small I}}=2.356\times10^5{D_{\rm Mpc}}^2S_{\rm H{\small I}}$, where $D_{\rm Mpc}$
is the distance to the galaxy in Mpc and $S_{\rm H{\small I}}$ is the total H{\sc i} flux
in Jy~km~s$^{-1}$. All H{\sc i} fluxes and hence H{\sc i} masses presented in Table 3
have been corrected for the primary beam response.

\begin{table*}
\centering
\begin{minipage}{170mm}
\caption{Summary of the H{\sc i} structures.\label{tbl-hidat}}
\begin{tabular}{llrccccl}
\hline\hline
Field  & Object & \multicolumn{1}{c}{H{\sc i} Peak Position}& $V_{\rm HI}$ & $S_{\rm HI}^{\rm~~~ a}$ 
       & log ($M_{\rm HI}/M_\odot$)${\rm^b}$ & $B_T^{\rm ~~c}$ & $M_{\rm HI}/L_B$\\
     & & \multicolumn{1}{c}{($^{h~m~s}$)~ ($^{\circ}~~~'~~~''$)~} 
       & (km~s$^{-1}$)
       & (Jy~km~s$^{-1}$)
       & 
       & (mag)
       & ($M_\odot/L_\odot$)\\
\hline
NGC~128 & UGC~298       & 00 29 51.315 +02 56 50.56 &3930 & 0.56 & 8.58 & 16.01 &~~ 0.25\\
        & NGC~127       & 00 29 13.765 +02 52 20.60 &4050 & 0.14 & 7.98 & 15.33 &~~ 0.03$^{\rm d}$\\
        & PGC~001760    & 00 28 37.717 +02 57 10.56 &4500 & 3.58 & 9.40 & 15.93 &~~ 1.27\\
\hline
NGC~3203 & AC dwarf I$^{\rm e}$ 
         & 10 19 26.984 $-$26 42 03.36 & 2560 &0.66 & 8.36 & 15.86 &~~ 0.22\\
\hline
NGC~7332 & SDSS~J223829  & 22 38 29.456 +23 51 30.80 & 1410 &0.32 & 7.59 & 16.94 &~~ 0.29\\
         & NGC~7339      & 22 37 48.308 +23 48 11.97 & 1300 &7.11 & 8.95 & 13.10 &~~ 0.19\\
         & SDSS~J223627  & 22 36 27.737 +23 43 00.44 & 1400 &0.59 & 7.87 & 15.74 &~~ 0.18\\
\hline
\end{tabular}
\tiny\
\noindent$^{\rm a}${All H{\sc i} fluxes have been corrected for the primary beam response.}\\
\noindent$^{\rm b}${Log of H{\sc i} mass in $M_\odot$ calculated adopting
a distance of 53.9, 38.2 and 23.0 Mpc for the first three galaxies, AC dwarf I, and the last
three galaxies, respectively. The distance to NGC~7332 is taken from the surface brightness
fluctuation measurements by \citet{tonry01}.}\\
\noindent$^{\rm c}${Extracted by \citet{leda00} from the Digitized Sky Survey (DSS). For 
galaxies not catalogued by \citet{leda00}, we determined the total $B$ magnitude from the DSS
image. }\\
\noindent$^{\rm d}${It is assumed that the H{\sc i} is associated only with NGC~127.}\\
\noindent$^{\rm e}${Object never been catalogued at
any wavelength before, hence it is named temporarily in this study.}
\end{minipage}
\end{table*}

\section{Results}
\label{result}
\subsection{NGC~128}
We find 9.6$\times10^7~M_\odot$ of H{\sc i} associated with the stellar disc of 
NGC~128. As shown on the right panel of Figure~\ref{fig-n0128hi}, the H{\sc i} is found 
only on one half of the disc, covering almost the entire stellar disc of its
neighbour, NGC~127. The H{\sc i} spatial extent is $\approx$1~arcmin, roughly 15~kpc 
at this distance. Although the peak intensity occurs between the
two galaxies,
it is closer to the systemic velocity of
NGC~127 (4094~km~s$^{-1}$, see Fig.~\ref{fig-n0128gp}). 
The H{\sc i} morphology and the velocity profile strongly suggest that the H{\sc i}
was originally bound to NGC~127 rather than NGC~128. In fact, the H{\sc i} feature is also 
seen in the Arecibo observations of \citet{bb77}. Their spectrum is comparable to that of the
VLA in its peak intensity and velocity width. However, these authors did not make
the connection with NGC~127.

We find two other H{\sc i} structures in this field, both nicely
coinciding with optically catalogued galaxies: UGC~298 and PGC~001760, as shown 
on the left panel of Figure~\ref{fig-n0128hi}. There is
no optical redshift measurement for UGC~298 to date, so the velocity from
our H{\sc i} study is the first. Adopting the same distance as NGC~128, the
H{\sc i} mass of UGC~298 is 3.8$\times10^8~M_\odot$. The redshift of 
PGC~001760 measured with our data is consistent with its
optical redshift (4427$\pm$50 km~s$^{-1}$; RC3). Adopting the same distance as
NGC~128, the H{\sc i} mass of PGC~001760 is 2.5$\times10^9~M_\odot$.
The total H{\sc i} line profile of the three H{\sc i} structures found 
in the NGC~128 field are shown in Figure~\ref{fig-n0128gp}
on the same scale.


\subsection{NGC~3203}
We find an H{\sc i} blob toward the southwest edge of NGC~3203's optical disc.
The H{\sc i} gas mass is 2.3$\times10^8~M_\odot$ at the distance of NGC~3203.
As shown in Figure~\ref{fig-n3203hi}, there is a dwarf galaxy
at the position of this H{\sc i} cloud, and the closest projected
distance that can be measured between the two galaxies (from edge to edge) is about 2.5~kpc.
This small $disky$ galaxy had not been optically-identified to date, and hence 
its redshift is not available
in any database. Although the centre of the H{\sc i} cloud and the optical centre
of the dwarf are offset by $\approx1~$kpc (assuming the same distance as
NGC~3203), they are largely coincident and the H{\sc i} cloud is elongated 
in the same direction as the major axis of the dwarf (almost perpendicular to NGC~3203's disc).
This is highly suggestive that the origin of the H{\sc i} cloud is the dwarf rather
than NGC~3203.
As shown in Figure~\ref{fig-n3203gp}, the H{\sc i} cloud velocity is found 
well within the range of velocities covered by the disc of NGC~3203 \citep[2424 $\pm~150~$km~s$^{-1}$;][]{cb04}. 
This makes the connection of the H{\sc i} cloud and the counter-rotating
ionized gas in NGC~3203 even more likely. The offset between the H{\sc i} peak
and the optical centre of the dwarf could well be an indication 
of the tidal interaction between the two galaxies. 

We do not find any other H{\sc i} gas in this field and name the newly discovered H{\sc i}-rich dwarf
``AC dwarf I''.

\subsection{NGC~7332}
We find 8.9$\times10^8~M_\odot$ of H{\sc i} in NGC~7332's neighbour, NGC~7339,
while we do not find any H{\sc i} gas directly associated with NGC~7332 itself.
Note, however, that the prior H{\sc i} detection of NGC~7332 by 
\citet{knapp78}, which was incorporated into the RC3 is likely to have come from 
Arecibo side-lobes 
picking up NGC~7339 as already suggested by others such as \citet{haynes81}.
The optical centres of the two galaxies are $\approx30~$kpc apart, but
the closest projected distance that can be measured is only $\approx$16~kpc
at the distance of the pair, as measured from the DSS2 image
down to the survey surface brightness limit. 

\begin{figure*}
\includegraphics[height=5cm]{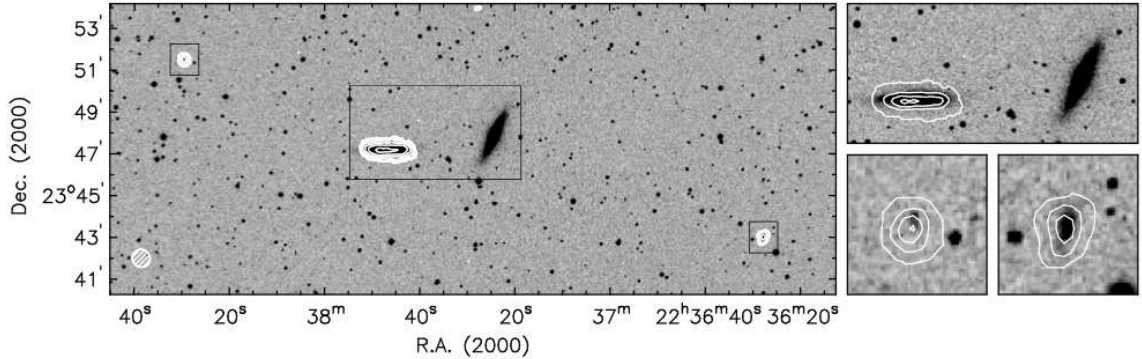}
\caption[]{On the left, the H{\sc i} found in NGC~7332's field is
shown in white contours overlaid on the DSS2 blue image. This H{\sc i} 
image was produced using the data cube smoothed to a beam of
$31''\times29''$ and a velocity resolution of $15~$km~s$^{-1}$, 
comparable to those of Morganti et al.'s (2006) WSRT data.
The contours correspond to H{\sc i} column densities of 1, 2.5, 5, 10, 25, 
...$\times10^{19}$~cm$^{-2}$. The smoothed synthesized beam 
is shown at the bottom left. The regions outlined with boxes are
magnified on the right. On the top right, the pair NGC~7332 (right) and NGC~7339 (left) 
is magnified (9$'\times4.5'$). The contours of the H{\sc i} gas are the same as the figure
on the left side. On the bottom right, 
2.5$'\times2.5'$ areas around the peaks of the two small H{\sc i} structures are
shown, again with the same contours.
These two galaxies are catalogued in the Sloan Digital Sky Survey (SDSS) but no optical
spectrum is available, and hence the optical redshifts are not known.\label{fig-n7332hi}}
\end{figure*}

\begin{figure*}
\includegraphics[bb=40 195 560 380, height=5cm]{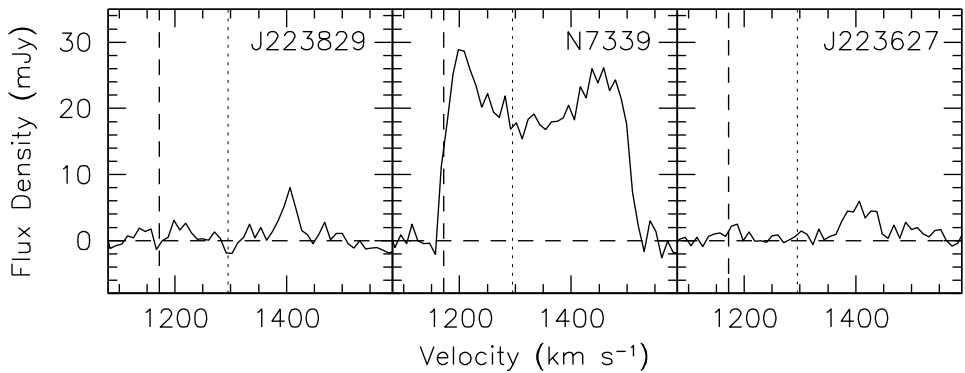}
\caption[]{Total H{\sc i} line profiles of the gas structures shown in 
Figure~\ref{fig-n7332hi}. 
The dashed line represents the optical velocity of NGC~7332; the dotted lines
represent its rotational velocity range ($\approx250$ km~s$^{-1}$), taken 
from \citet{fb04}. \label{fig-n7332gp}}
\end{figure*}

\citet{morganti06} reported an H{\sc i} cloud of 
$6\times10^6~M_\odot$ at a redshift of 1250~km~s$^{-1}$ from Westerbork Synthesis Radio Telescope (WSRT) 
data, almost halfway 
spatially between the two galaxies. To confirm this, we have tapered our
VLA data to $31''\times29''$, a beam similar to that of the WSRT ($41''\times30''$),
and Hanning smoothed the cube to $15$~km~s$^{-1}$, a comparable velocity resolution
to that of the WSRT cube ($16~$km~s$^{-1}$). 
We present in Figure~\ref{fig-n7332hi} the H{\sc i} image obtained from the smoothed
and tapered cube using the same contour levels as those of \citet{morganti06}.
Although our 3$\sigma$ H{\sc i} mass sensitivity in this cube is $1.2\times10^6~M_\odot$,
we do not detect the $6\times10^6~M_\odot$ H{\sc i} cloud seen by \citet{morganti06}.




More recently, \citet{minchin10} have discovered a large H{\sc i}
structure covering both NGC~7339 and NGC~7332, with the H{\sc i} peak 
coinciding with the center of NGC~7339. As shown in their H{\sc i}
image and renzogram (their Figs. 8 and 10), this large H{\sc i} envelope
appears to be an extension of NGC~7339's H{\sc i} disc. 
\citet{morganti06}'s cloud could well be the part of this feature, 
and it is possible that they detected one of the high density regions.
Indeed, the total VLA H{\sc i} flux is $\approx$20\%
lower than the single-dish flux of \citet{minchin10}, and it is likely 
that we are missing some diffuse H{\sc i}.

\begin{figure}
\begin{center}
\includegraphics[width=7.5cm]{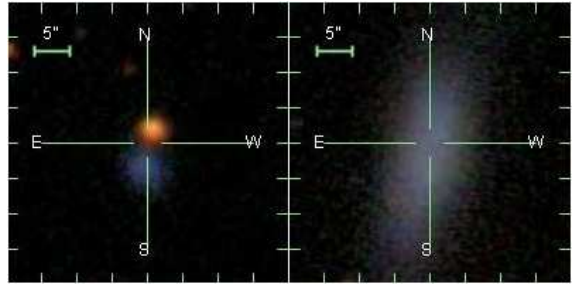}
\end{center}
\caption[]{Sloan Digital Sky Survey (SDSS) Data Release (DR) 7 colour images 
of the two small H{\sc i}-rich galaxies in NGC~7332's field. The H{\sc i} peak 
of J223829 in Figure~\ref{fig-n7332gp} coincides well with the optical peak of 
the blue object shown on the left. The red object is likely to be a foreground star. 
\label{fig-n7332dw}}
\end{figure}

We also find two other H{\sc i} clouds in the cube at a similar redshift
($\approx$1400~km~s$^{-1}$), as shown in Figures~\ref{fig-n7332hi} and \ref{fig-n7332gp}.
These two H{\sc i} structures coincide with dwarf galaxies
identified in the Sloan Digital Sky Survey (SDSS)
(Fig.~\ref{fig-n7332dw}), but their optical redshifts are unknown. At the 
location of the H{\sc i} structure to the southwest of NGC~7332, there is only
one possible optical counterpart, J223627.84$+$234257.5, while
there are two candidates at the location of the northeast H{\sc i} structure,
under a single catalogue number, J223829.69$+$235131.2. Based on
its colour and fuzziness, the blue object is more likely to be the true 
counterpart of the H{\sc i} gas. The red object shows no substructure
but a similar colour as other faint stars in the field, hence it
is probably a faint foreground star. In fact, the H{\sc i} 
peak coincides better with the optical centre of the blue galaxy. 
The two H{\sc i}-rich galaxies are located almost at the half power
point at the primary beam and their H{\sc i} fluxes measured after primary beam correction are
consistent within the errors with those of \citet{minchin10}. 
The H{\sc i} masses are about $3.9\times10^7~M_\odot$
and $7.4\times10^7~M_\odot$, respectively.


Unlike for NGC~128 and NGC~3203, we do not find any direct evidence for 
interaction between NGC~7332 and its neighbours, NGC~7339 and the two H{\sc i}-rich 
dwarfs.
However, the H{\sc i} gas morphology and kinematics found by \citet{minchin10}
are highly suggestive that NGC~7332 and NGC~7339 are tidally affecting each other,
even considering the large beam of the single dish. 
It is also noteworthy that NGC~7332 is in a gas rich environment,
with several potential gas donors. 

\section{Discussion}
\label{discuss}
\subsection{The origin of counter-rotating gas}
In this H{\sc i} follow-up study, we have found H{\sc i} gas in the disc of
two out of the three galaxies hosting kinematically-decoupled ionized gas observed.
In both cases, the H{\sc i} gas is highly offset from the optical centre of
the counter-rotator, being found only on one side of the disc. In addition, in
both cases, the H{\sc i} gas covers a large fraction of a dwarf companion. In the
case of NGC~127, the companion of NGC~128, the H{\sc i} covers most of its
stellar body, although it is spatially offset and more extended toward
NGC~128, filling up the intergalactic space between the pair
where diffuse light (likely tidally-stripped stars) is present.
The centre of the H{\sc i} velocity coincides well with the optical
systemic velocity of NGC~127, and the entire H{\sc i} velocity range is well
within the range covered by NGC~128's rotational velocity. In the case of the dwarf
companion of NGC~3203, AC dwarf I, the H{\sc i} covers its entire stellar body.
AC dwarf I appears to be disky, and the H{\sc i} cloud is enlongated 
in the same direction as its disc, the major axis being perpendicular
to that of NGC~3203. The H{\sc i} velocity range overlaps with the rotational
velocity range of NGC~3203 by a few tens of km~s$^{-1}$. 

These observations strongly suggest gas accretion onto NGC~128
from NGC~127, and onto NGC~3203 from AC~dwarf~I. 
The counter-rotating ionized gas present in NGC~128 and NGC~3203 
\citep[see Fig. 1 of][]{bc06} is thus likely to originate from this
externally accreted gas. As the ISM from 
the dwarf neighbours enters in retrograde motion with respect to the
stellar discs, it must be shock-heated and is (partially)
currently observed in the form of ionized gas. 

In NGC~7332, we have not found any H{\sc i} gas directly associated with the disc
in our VLA observations.
\citet{minchin10} however find a large scale H{\sc i} envelope centered on its neighbour,
NGC~7339, that extends all the way to NGC~7332, and it covers its entire stellar disc.
In addition, both the H{\sc i} and the optical morphologies
show an asymmetry in the outer disc while the inner disc appears quite symmetric. Our 
high-resolution VLA data reveal a short tail pointing to the north east and the
stellar envelope is more extended and thicker on the west side, towards NGC~7332. 

These facts strongly suggest a tidal interaction between NGC~7332 and NGC~7339.
If the H{\sc i} in NGC~7339 was more extended than the stellar disk, as is often seen in 
spirals, then a tidal interaction would have stripped the gas before the stars are pulled out.
The stripped gas would likely have fallen toward the center of NGC~7332, forming the
counter-rotating gas in its centre.
This is also supported by the kinematics of the counter-rotating ionized gas observed by 
\citet{fb04}, which is tilted with respect to the equatorial plane in the direction
of NGC~7339. 

\begin{table*}
\centering
\begin{minipage}{142mm}
\caption{Dynamical timescales for the neighbours.\label{tbl-timescale}}
\begin{tabular}{llrrrrrrr}
\hline \hline
Field &
Companion  & 
$V_{\rm rot}^{~~~\rm a}$~  &
$R_{\rm disc}^{~~~~\rm b}$ &
$M_{\rm comp}^{~~~~~~\rm c}$ &
$d_{\rm sep}^{~~~~\rm d}$  &
$\Delta V$~~~ &
$t_{\rm fric}$~~ &
$t_{\rm orbit}$~~ \\
& &
(km~s$^{-1}$)&
(kpc) &
($10^9~M_\odot$) &
(kpc)  &
(km~s$^{-1}$)&
($10^9$ yr) &
($10^9$ yr) \\
\hline
NGC~128 & UGC~298  & 107~~~ & 12~~~ & 31.7~~~ & 162~ & 399~~~ & 12.0~~ & 3.3~~~~\\
        & NGC~127  & 125~~~ & 11~~~ & 39.8~~~ &  12~ & 150~~~ &  0.3~~ & 0.5~~~~\\
     & PGC~001760  & 134~~~ & 20~~~ & 83.0~~~ & 162~ & 300~~~ &  5.8~~ & 3.3~~~~\\
\hline
NGC~3203&AC Dwarf I&  47~~~ & 11~~~ &  4.9~~~ &  14~ & 150~~~ &  0.5~~ & 0.5~~~~\\
\hline
NGC~7332 & J223829 &  37~~~ &  5~~~ &  1.6~~~ & 102~ & 225~~~ &  49.3~~ & 2.8~~~~\\
         &NGC~7339 & 172~~~ & 22~~~ &151.4~~~ &  31~ & 140~~~ &  -2.7~~ & 1.4~~~~\\
         & J223627 &  37~~~ &  5~~~ &  1.6~~~ &  92~ & 230~~~ &  41.3~~ & 2.5~~~~\\
\hline
\end{tabular}
\tiny
$^{\rm a}${Disc rotational velocities of companions estimated using $W_{\rm HI}/2$. The inclination
w.r.t. the observer's line-of-sight has not been taken into account, hence our estimations are
lower limits for $V_{\rm rot}$.}\\
$^{\rm b}${The disc size has been measured
using the H{\sc i} extent, except for NGC~127 which is larger in the optical. The same
distance as their primary galaxies have been adopted.}\\
$^{\rm c}${Dynamical masses of companions calculated using $V_{\rm rot}$ and $R_{\rm disc}$ assuming a rotationally supported
disc and spherical symmetry, i.e. $M_{\rm comp}=V_{\rm rot}^2R_{\rm disc}/G$.}\\ 
$^{\rm d}${Separations of the pair centers.}\\
\end{minipage}
\end{table*}

\subsection{The fate of companions}
The time required for a companion to spiral into the primary galaxy
can be characterized by the dynamical friction timescale \citep{bt87}, 
\begin{equation}
t_{\rm fric}=\frac{1.17}{\rm ln \Lambda}\frac{r_{\rm i}^2 v_{\rm c}}{GM},
\end{equation}
where $r_{\rm i}$ is the initial radius of the orbit, $v_{\rm c}$ is the orbital velocity, $G$ is
the gravitational constant, $M$ is the mass of the companion, and ln$\Lambda$ is the
Coulomb logarithm. As described in \citet{bt87}, the appropriate $\Lambda$ can be approximated 
by $r_{\rm i} v_{\rm c}^2/GM$ for this case. In more practical units, this relation can be rewritten as
\begin{equation}
\frac{t_{\rm fric}}{\rm yr}
\approx\frac{2.26\times10^{14}}{\rm ln \Lambda}
\left( \frac{r_{\rm i}}{\rm kpc} \right)^2
\left( \frac{v_{\rm c}}{\rm km~s^{-1}} \right)
\left( \frac{M}{M_\odot} \right)^{-1}.
\end{equation}

We adopt the projected distance from the primary
galaxy to its companion ($d_{\rm sep}$) and their velocity difference ($\Delta V$) 
as the orbital radius $r_{\rm i}$ and velocity $v_{\rm c}$, respectively. The same line-of-sight
distances as the sample galaxies have been assumed. 
Companion masses have been estimated 
using the disc size ($R_{\rm disc}$) and the H{\sc i} linewidth measured
at 20\% of the peak flux $W_{\rm HI}$, assuming that the companions are rotationally supported 
($M_{\rm comp}\approx V_{\rm rot}^{~2}R_{\rm disc}/G$, where $V_{\rm rot}\equiv W_{\rm HI}/2$). 
The disc size has been estimated from the H{\sc i} extent except for NGC~127 which has a larger
stellar disc compared to its H{\sc i} disc.
The dynamical friction and orbital timescales of the neighbours found with H{\sc i} gas 
are compared in Table~\ref{tbl-timescale}. 

The orbital
period around the galaxy hosting counter-rotating gas has been estimated for each neighbour by
$t_{\rm orbit}\approx 2\pi d_{\rm sep}/\Delta V$.  
For NGC~128 and NGC~3203, $t_{\rm fric}$ and $t_{\rm orbit}$ of their nearest companions are
comparable and of the order of 10$^8$~yr. This implies that these pairs are likely to soon merge and
form a single system, fueling the primary galaxies with $10^7-10^8~M_\odot$ of H{\sc i} gas.
In fact, the H{\sc i} discs of NGC~127 and AC~Dwarf~I appear to cover some of their primary 
galaxy's discs already. This strongly supports the hypothesis that these two neighbours
have been feeding NGC~128 and NGC~3203 with gas. For UGC~001760, 
$t_{\rm fric}$ and $t_{\rm orbit}$ are comparable but it is expected to take much longer than
10$^9$~yr to merge with NGC~128, while it will take even longer for
UGC~298 to spiral into NGC~128.

The estimated Coulomb logarithm of NGC~7339 is negative (i.e. $v_c^2/r_i<GM/r_i^2$), 
suggesting that this galaxy is already in the process of merging with
NGC~7332.
This is also supported by the observations of \citet{minchin10}.
We also have estimated $t_{\rm fric}$ and $t_{\rm orbit}$  for the two farther 
dwarf neighbours, using $W_{20}$ and assuming that they are rotationally supported.
For these two dwarfs, $t_{\rm fric}$ measured using the H{\sc i} extent is about
an order of magnitude larger than $t_{\rm orbit}$, and
it may take many orbital periods for these two dwarf galaxies to merge. 
As they approach NGC~7332 and 7339, however, it is still possible that
diffuse gas outside their stellar body at large radii could be stripped by the halo of
their bigger neighbours, also fueling the pair \citep[as suggested by][for 
satellites of the Milky Way]{gp09}. 

\subsection{Incidence of gas counter-rotation and frequency of accretion from 
cold gas blobs}

The fraction of all S0 galaxies with ionized gas counter-rotation is found
to be some 15 per cent and slightly increases ($\approx23$ per cent) 
for galaxies with detected ionized gas only \citep[][and references therein]{bc06}.
This provides a lower limit to the gas accretion frequency among
lenticular galaxies. 
This fraction however does not include 1) the elliptical galaxy population
and 2) gas accreted in prograde orbit. Taking these into account, the
inferred frequency of gas accretion in early-type galaxies overall is 
expected
to increase. This is supported by the high H{\sc i} detection rate 
found by \citet[][see also Oosterloo et al. 2010]{morganti06} in their carefully selected E/S0 galaxies,
many of which show direct evidence or hints of interaction. All these
facts suggest that gas accretion from cold gas blobs is common
among E/S0 galaxies and is thus an important  source for their continous 
build-up.

In gas-rich late-type galaxies, where the gas acquired from outside 
(either prograde or retrograde) is likely to be swept away by existing
gas on relatively short timescales, evidence of gas accretion (especially
from small gas blobs) may not be detected as frequently as in the E/S0 
population \citep{bc06}. Assuming the same statistics as for
early-type galaxies, however, merging of small gas-rich dwarfs is also expected 
to be common among gas-rich late-type galaxies, and also may play an
important role in replenishing their discs. 

\subsection{Gas accretion and galaxy evolution}
It is worth noting that all three counter-rotators targeted here are gas-poor, barred S0 
systems. It is unclear, however, whether gas accretion has played any role in exhausting 
the original gas in the host galaxy (e.g. by triggering a burst of star formation)
or whether gas counter-rotation is preferentially observed in early-type systems simply 
because retrograde gas accretion can last longer than in gas-rich galaxies \citep[where it is 
likely to be swept away by co-rotating gas; see][]{wk86,bc06}.

In the former case, the recent star formation rate is expected to be quite high, assuming
spirals with normal gas contents are the ancestors of the hosts. As discussed in
detail in the previous sections, the time the nearest neighbours will take to 
merge into the hosts is of order a few $10^8$~yrs. Assuming a $L^*$
host with $M_{\rm gas}\approx5\times10^9~M_\odot$, our estimated dynamical friction
timescales imply that these counter-rotators should have been forming stars 
at a rate of 5--50~$M_\odot~$yr$^{-1}$ to use up their original gas.
We however do not find any signature of such high star formation rates in the sample hosts, 
based on their low infrared fluxes and ionized gas emission \citep[][]{iras89,bf99,cb04}.
In addition, it is unlikely that only the gas in the host galaxies would be
exhausted but not the gas in their companions, particularly in 
cases like NGC~127 and AC dwarf I, where we are witnessing
the gas being captured by the primary galaxies. 

Hence we conclude
that gas accretion (and the subsequent starburst) is not likely to be responsible for the early-type morphology of our
counter-rotators. 
It is however still possible that tidal interactions, along with gas accretion, played a role
in forming the bar and bulge in these galaxies by 1) inducing instability in the disc, 
then 2) feeding the centre with the accreted gas, possibly triggering a nuclear starburst
in some cases \citep[e.g.][]{ce93}.


\section{Conclusion}
In this H{\sc i} follow-up study of three galaxies where ionized gas was
found to be rotating in the opposite direction to the stars
in the central few kpc, we systematically find evidence for cold gas 
accretion from
neighbours, likely to be the origin of the counter-rotating
ionized gas. The dynamical friction timescales of the neighbours
are short and comparable to their orbital periods around the counter-rotators,
suggesting that the pairs will merge and form a single system within about a gigayear,
fueling the primary galaxy with $10^7-10^8~M_\odot$ of H{\sc i}. 
We do not find strong evidence that gas accretion is directly responsible 
for the early-type morphology of the sample hosts, i.e. the transformation 
from gas-rich spirals to gas-poor lenticular systems. However, it could well have
helped to form and fuel the bar in these galaxies.

Our data, along with statistics from the literature \citep[e.g.][]{bc06},
suggest that the accretion of cold gas blobs is an important source of gas in galaxies,
and hence helps them continue to grow. 
Although the accretion of gas blobs is expected to be
more easily detected in gas-poor early-type systems, assuming the same
frequency for late-type galaxies implies that cold accretion
can also play a significant role in replenishing the disc
of spirals.

\section*{Acknowledgments}
We are grateful to the referee, Dr. R. Minchin for his useful comments which were helpful to improve the manuscript.
This work has been supported by National Research Foundation of Korea grant 2011-8-0993, Yonsei research 
grant 2010-1-0200, a Yonsei Global research grant and NSF grant AST-00-98249 to Columbia University.
Support for this work was also provided by the National Research Foundation of Korea to the Center for Galaxy Evolution Research.
Funding for SDSS and SDSS-II has been provided by the Alfred P. Sloan Foundation, the 
Participating Institutions, the National Science Foundation, the U.S. 
Department of Energy, the National Aeronautics and Space Administration, 
the Japanese Monbukagakusho, the Max Planck Society, and the Higher 
Education Funding Council for England. The SDSS web site is 
$http://www.sdss.org.$



\end{document}